# A Recursive Definition of Goodness of Space for Bridging the Concepts of Space and Place for Sustainability

Bin Jiang

Faculty of Engineering and Sustainable Development, Division of GIScience
University of Gävle, SE-801 76 Gävle, Sweden
Email: bin.jiang@hig.se

*(Draft: July 2018, Revision: September, November 2018, February, July 2019)*

*In a good system, we would expect to find the following conditions: Any identifiable subsystems, we would hope, would be well – that is to say, in good condition. And we would hope that the larger world outside the complex system is also in good order, and well. Thus, the mark of a good system would be that it helps both the systems around it and those which it contains. And the goodness and helping towards goodness is, in our ideal complex system, also reciprocal. That is, our good system, will turn out to be not only helping other systems to become good, but also, in turn, helped by the goodness of the larger systems around it and by the goodness of the smaller ones which it contains.*

Christopher Alexander (2003, p. 6)

**Abstract**
Conceived and developed by Christopher Alexander through his life's work: *The Nature of Order*, wholeness is defined as a mathematical structure of physical space in our surroundings. Yet, there was no mathematics, as Alexander admitted then, that was powerful enough to capture his notion of wholeness. Recently, a mathematical model of wholeness, together with its topological representation, has been developed that is capable of addressing not only why a space is good, but also how much goodness the space has. This paper develops a structural perspective on goodness of space – both large- and small-scale – in order to bridge two basic concepts of space and place through the very concept of wholeness. The wholeness provides a *de facto* recursive definition of goodness of space from a holistic and organic point of view. A space is good, genuinely and objectively, if its adjacent spaces are good, the larger space to which it belongs is good, and what is contained in the space is also good. Eventually, goodness of space – sustainability of space – is considered a matter of fact rather than of opinion under the new view of space: space is neither lifeless nor neutral, but a living structure capable of being more living or less living, or more sustainable or less sustainable. Under the new view of space, geography or architecture will become part of complexity science, not only for understanding complexity, but also for making and remaking complex or living structures.

## 1. Introduction
What is considered to be a good space? This is the primary question that motivates this paper. A good space – for example traditional and vernacular buildings and cities (c.f. Figure 2 for some examples) – is one in which or with which people have good feelings, in terms of belonging, healing or well-being. The sense of well-being is a deeper, psychological and emotional comfort, in which people could feel their own existence in space rather than thermal comfort, or energy saving, or comfort of illumination on surfaces (Alexander 2016). On the contrary, a space that fails to create such positive feelings is considered to be a bad space, for example many of modernist buildings and cities (Mehaffy and Salingaros 2006, Salingaros 2006, Curl 2018). This human experienced sense of space has another name called place in the literature (e.g. Tuan 1977, Goodchild and Li 2011, Mennis and Yoo 2018); see a further discussion in Section 5. The notion of space, in the context of this paper, refers not only to large-scale spaces such as countries, cities, campus and buildings, but also small-scale spaces such as ornaments, artifacts, and simple shapes. Every year, numerous people from all over the world spend large amounts of money traveling to good places/locations for vacations, conferences, and other activities. There is little doubt that the Earth's surface is a good space. However, the goodness varies from one location to another and can also change over time (for example, a space can become a better one). While urban professionals and architects are responsible for planning, designing, and making sustainable cities and buildings, it is geographers that study the Earth's surface. Collectively, geographers, urban professionals, and architects aim to make better and sustainable built environments. As a note, instead of the mainstream technology-oriented sustainability such as green or energy saving buildings, this paper adopts the notion of sustainability based on morphogenesis, as advocated by Alexander (2004). Given the circumstance, the traditional pre-20$^{th}$ century cities such as Amsterdam, Rome and Venice are considered to be sustainable rather than the modernist cities built over the past 100 years or so. This is the reason that in this paper goodness of space holds the same meaning of sustainability of space, and both can be interchangeably used.

What, then, is the standard or criterion for goodness of space? It is conventionally considered to be based on opinions or personal preferences. Christopher Alexander (2002–2005), however, challenged this view, arguing that 90% of our feelings – actually feelings arising out of wholeness or feeling of wholeness – are shared, and there is a shared notion of goodness of space regardless of people's faiths, ethics, and cultures. Goodness of space is structural – being objective – and it can be well assessed using the concept of wholeness, which is defined as a recursive structure. Simply put, a space is good, genuinely and objectively, if its adjacent spaces are good, the larger space to which it belongs is good, and what is contained in the space is also good (Alexander 2002–2005). This is essentially a recursive definition of goodness of space. In other words, goodness of a space is not determined by the space itself, but rather iteratively or recursively determined by its neighboring spaces, by the larger space to which it belongs, and by the smaller spaces it contains. The central argument of this paper is that goodness of space is a matter of fact rather than of opinion or personal preference.

This argument is founded upon the concept of wholeness. The wholeness is to goodness of space what temperature is to warmness or coldness. The feeling of warmness or coldness is largely objective, governed by temperature. This is the same for the feeling of goodness of space, governed by wholeness, which is the central theme of the four-volume book *The Nature of Order* (Alexander 2002–2005) (in the literature and by Alexander himself, the four volumes are actually four independent books). A space with a high degree of wholeness is also called living structure. Unlike wholeness in Gestalt psychology (Koffka 1936) yet very much like that in quantum physics (Bohm 1980), Alexander's wholeness is defined as a describable mathematical structure that exists pervasively in physical space and is reflected psychologically in our minds. Although mathematical principles of wholeness were fully explored and discussed (Alexander 2002–2005, Volume 1), there was no mathematics – as Alexander admitted then – that was powerful enough to capture wholeness. Due to the lack of mathematics, Alexander (2002–2005) used over 2000 pictures, photos, and drawings to present his design ideas, thoughts and theories. We believe that the lack of mathematics has made his profound thoughts and theories hard to be well understood and applied.

A mathematical model of wholeness has been recently developed, and it is capable of capturing Alexander's initial definition of wholeness (Jiang 2015c). Yet this mathematical model cannot be directly applied to current geographic information systems (GIS) representations that are largely made of geometric primitives such as pixels, points, lines, and polygons (Bian 2007). These representations – still called topology in the current GIS literature, but geometric in essence – concentrate on geometric primitives of pixels, points, lines, and polygons, and details of locations, sizes, and directions, which are framed under the mechanistic world view of Descartes (1637, 1954) through Newtonian absolute and Leibnizian relative views of space (Belknd 2013). These geometric primitives are more or less similar in size as individualized mechanical pieces, rather than a coherent whole that commonly consists of far more small things than large ones. In other words, the wholeness or living structure is not only recursive, but also composed of far more smalls than larges, a distinctive nature of geographic space or phenomena (Jiang 2015b, 2015c, Christaller 1933, 1966, Zipf 1949). Thus, conventional GIS representations – or Cartesian mechanistic thinking in general – failed in capturing the underlying wholeness of geographic space. Instead, a topological representation among geometrically coherent and meaningful entities, such as streets and cities rather than line segments and polygons, is able to capture the underlying living structure of geographic space. The topological representation that concentrates on overall character rather than geometric details of locations, sizes, and directions is profoundly important to a better understanding of the two concepts of space and place.

This paper presents a recursive definition of goodness of space, in order to bridge together two basic concepts of space and place, and formalize them through the very concept of wholeness. We argued, as Alexander did, that goodness of space is a matter of fact rather than opinion or personal preferences. The formalization may significantly contribute to the ongoing effort (e.g. Goodchild 2004, Sui 2004) to explore fundamental laws of geography and/or geographic information science. A space is considered to be a living structure governed by the wholeness, being external, whereas the human experienced sense of space, being internal, arises directly from the living structure or wholeness. The three-folder contribution of this paper is more philosophical than empirical (which was cited when appropriate): (1) argues that mechanistic thinking in general – or mechanistic representation of GIS in particular – failed to recognize living nature of geographic space (see this Introduction briefly, yet more details in Jiang and Ren (2018)), (2) illustrates that space and place can be bridged together thought wholeness, rather than separated as currently conceived; and (3) demonstrates that goodness of space is a matter of fact rather than opinion or personal preference.

We develop our discussion in three stages. First, we present in Section 2 two fundamental laws of geographic space or living structure in general – scaling law and Tobler's law – and in Section 3 wholeness as a mathematical structure of physical space, and the 15 properties of wholeness in relation to the two laws and two design principles of differentiation and adaptation. Next, we demonstrate in Section 4 why goodness of space is a matter of fact rather than opinion for both small- and large-scale spaces. Finally, we discuss further in Section 5 on the implications of the recursive definition of goodness of space on geography and for sustainability, before draw in Section 6 a conclusion, pointing out how geography or architecture can be significantly reshaped towards a major science if wholeness is adopted as its scientific foundation.

**2. Two fundamental laws of living structure: Scaling law and Tobler's law**
A space with a high degree of wholeness (or life or goodness or sustainability) is called a living structure. The kind of living structure is governed by two fundamental laws – scaling law and Tobler's law – that are mutually complementary (Table 1, Jiang 2018b). Scaling law states that there are far more small things than large ones across all scales ranging from the largest to the smallest (Jiang 2015a). It should be noted that scaling law is very much relaxed; unlike the universal rule (Salingaros and West 1999) that is based on power laws, scaling law could imply power laws, lognormal, exponential functions or even skewed normal distributions as long as the notion of far more smalls than larges recurs across on at least two scales (see Figure 1 and the related discussions) rather than occurring on just one scale. In contrast to scaling law, Tobler's law (1970) – also widely known as the first law of geography – states that things are more or less similar on each scale (Note: the scale here means the size or hierarchical level rather than the map scale), or that nearby things tend to be more or less similar: *everything is related to everything else, but near things are more related than distant things* (Tobler 1970, p. 236).

These two laws reflect two spatial properties – spatial heterogeneity and spatial dependence – that are available in both space and time. For example, there are far more low housing prices than high housing prices on the city scale, yet on a given neighborhood scale housing prices tend to be more or less similar. Likewise, there are far more ordinary weather conditions than extraordinary ones, yet today's weather tends to be more or less similar to yesterday's weather. Scaling law reflects heterogeneity and inter-dependence across all scales globally, whereas Tobler's law reflects homogeneity and dependence on each of the scales locally. The spatial heterogeneity should be understood in the context of scaling law (Jiang 2015b), because all geographic features are fractal or scaling, given the first perspective and scope (Jiang and Yin 2014, Gao et al. 2017). Table 1 summarizes these mutually complementary aspects between these two laws.

Table 1: Comparison between scaling law and Tobler's law
(Note: These two laws complement each other and recur at different levels of scale in geographic space or the Earth's surface in general.)

| Scaling law | Tobler's law |
| --- | --- |
| far more small things than large ones | more or less similar things |
| across all scales | available on one scale |
| without an average scale (Pareto distribution) | with an average scale (Gauss distribution) |
| long tailed | short tailed |
| interdependence or spatial heterogeneity | spatial dependence or homogeneity |
| disproportion (80/20) | proportion (50/50) |
| complexity | simplicity |
| non-equilibrium | equilibrium |

To further illustrate these two laws, let us use two sets of 10 numbers: (a) 1, 1/2, 1/3, …, 1/10, and (b) 10, 9, 8, …, and 1. The first 10 numbers decrease nonlinearly through a power function, namely Zipf's law (1949) (Note: Zipf's law indicates statistical regularity, so it should not be expected to have any real-world data to be exactly on the power law curve.), whereas the second 10 numbers decrease linearly as an arithmetic series. The first 10 numbers are more pervasively seen in geography than the second 10 numbers are, because geographic space or phenomena – given the right perspective and scope – is essentially fractal rather than Euclidean, complex rather than simple, and nonlinear rather than linear. The first and second 10 numbers meet respectively scaling law and Tobler's law. The two sets differ dramatically (Figure 1). More differences between the two data sets can be explored through head/tail breaks (Jiang 2015a), which is a classification scheme and visualization tool for data with a heavy tailed distribution. The head/tail breaks is a de facto recursive function for deriving the underlying scaling hierarchy of data or pheneomana (see Recursive function 1). It should be noted that there is no need to examine whether data is long-tailed or short-tailed: simply applying head/tail breaks to your data, if the notion of far more smalls than larges recurs on at least twice, the data is long-tailed (see Figure 1a); otherwise it is short-tailed.

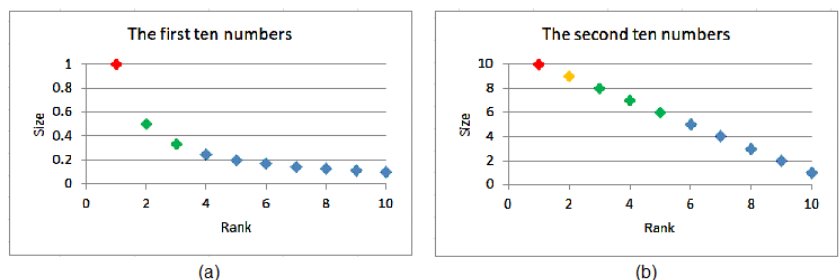

(a) (b)

Figure 1: (Color online) Rank-size plots for the two sets of 10 numbers
(Note: A rank-size plot, showing rank on the x-axis and size on the y-axis, differs from a histogram, in which size is shown on the x-axis and frequency on the y-axis. The scaling or notion of far more smalls than larges recurs twice, so there are three classes for the first 10 numbers, indicted by the three colors. However, the four colors in the right plot are not classes, but represent the four recurrences of the proportion between smalls and larges.)

Recursive function 1: Head/tail breaks 1.0
(Note: A 40% threshold is set for every head or iteration.)

```
Recursive function Head/tail Breaks:
    Rank the input data from the largest to the smallest
    Break the data into the head and the tail around the mean;
    // the head for those above the mean
    // the tail for those below the mean
```

```
        While (head <= 40%):
            Head/tail Breaks (head);
End Function
```

The head/tail breaks is further explained in detail through illustrations. For the first 10 numbers the mean is 0.29, which partitions the 10 numbers into two parts: the first three, as the head part (accounting for 30%), are greater than the mean, and the remaining seven (70%) as the tail are less than the mean. This completes the head/tail breaks on the first scale or iteration. On the next scale, for the three in the head, the mean is 0.61. This mean value further partitions the three largest numbers into two parts: 1 (33%), as the head, is greater than the mean 0.61, and 1/2 and 1/3 (67%) as the tail are less than the mean. At the end of the second iteration of head/tail breaks, the 10 numbers are divided into three classes: [1], [1/2, 1/3], [1/4, 1/5, …, 1/10] (Figure 1a), indicating seven smallest, one largest, and two in between the smallest and the largest (Note: head/tail breaks recursively progresses for the head part rather than the tail part); or put in general, numerous smallest, a very few largest, and some in between the smallest and the largest. Given a greater number of values with a long-tailed distribution, the head/tail breaks can be iterated in a higher number of scales (Jiang 2015a), rather than two scales in the above illustration.

As can be seen from the previous head/tail breaks process, the head percentage of every iteration or equivalently scale, is less than 40%, implying disproportionality between the head and the tail. However, this 40% threshold for the head is too restricted for many real-world data. Instead, head/tail breaks 2.0 below is a more relaxed version (see Recursive function 2).

Recursive function 2: Head/tail breaks 2.0
(Note: A 40% threshold is set not for every head, but for the average of all heads, shown as ~head.)

```
Recursive function Head/tail Breaks:
    Rank the input data from the largest to the smallest
    Break the data into the head and the tail around the mean;
    // the head for those above the mean
    // the tail for those below the mean
    While (~head <= 40%):
        Head/tail Breaks (head);
End Function
```

Unlike the first 10 numbers with a long-tailed distribution, the second 10 numbers are short tailed, so the head/tail breaks approach does not apply. To know why, let's try to conduct the head/tail breaks process for the second 10 numbers. The mean of the 10 number is 5.5, so it partitions the 10 numbers into two equal parts: five above the mean, and five below the mean, which violates the condition of small head and long tail. For the five numbers above the mean [10, 9, 8, 7, 6], their mean is 8, which again partition the numbers into two equal parts, which again violates the condition of head/tail breaks. In other words, on every iteration, the ratio between the head and the tail is equal proportion (e.g. 50/50 as indicated in Table 1). Figure 1b shows the number of times the equal proportion recurs rather than scaling as shown in Figure 1a.

## 3. Wholeness and its 15 properties bearing two fundamental laws and two design principles

In any space, numerous spatially coherent sets and subsets – or, equivalently, wholes and sub-wholes – that are nested within each other, with varying degrees of living structure for each, constitute a nested system of wholes or sub-wholes that cover the space. The nested system captures the kinds of deep structure of wholeness. The sub-wholes are termed as centers, which are the basic constituents of wholeness. These centers individually or as a coherent whole – holistically – possess many of fifteen properties (Figure 2) that are pervasively seen in our surroundings, not only in nature, but also in what we build or make (Alexander 2002–2005). Interested readers should refer to Alexander (2002–2005, p. 143–298) in which over 200 photos and pictures – some of which are shown in Figure 2 – are provided for accounting for these profound properties. This section provides a brief introduction to wholeness, its 15 properties, and their underlying laws: scaling law and Tobler's law and design principles: differentiation and adaptation.

The first property of levels of space is really about scaling law, recurring of far more smalls than larges. As a law, it governs the process of inducing more living centers, being implicit in many other properties, e.g., the void. Many of the 15 properties are used to create more living centers, such as thick boundaries, alternating repetition, local symmetries, contrast, and gradients. Note that if the created centers are at a same level, they must meet Tobler's law; if across different levels, they must meet scaling law. The property of simplicity and inner calm is about something on one scale, so Tobler's law and adaptation principle apply. Nearly all of the 15 properties are about differentiation, that is, to create more living centers, through the differentiation principle. In the course of differentiation, newly created centers must be adapted each other to be within the coherent whole or living structure. We will come back to this point again later in Figure 5.

The 15 properties, largely used for transforming centers, can be summarized by two laws – scaling law and Tobler's law (as discussed in Section 2 above) – and two design principles of differentiation and adaptation. These two fundamental laws are largely statistical regularities that underlie the 15 properties of wholeness or living structure. The more 15 properties, the higher the degree of wholeness. Thus, the 15 properties are also called transformation properties. The differentiation principle means that a space is continuously or timelessly (Alexander 1979) differentiated to induce far more small centers than large ones (c.f. Figure 3 for an illustration), governed by scaling law. The adaptation principle implies that nearby centers should be adapted each other, governed by Tobler's law. As time goes, a space is recursively transformed and results in many nested and overlapped living centers. A space is capable of becoming more living, if the wholeness-enhancing transformation is applied. "*We have a vision, now, of buildings taking their form continuously through a smooth step-by-step process in which each step preserves the structure of what was there before*", as remarked by Alexander (2002–2005, Volume 3, p. 678). This is also the vision of the timeless way of building Alexander (1979) had 25 years back at the time of pattern language, and eventually the theory of wholeness – the 15 transformation properties, or the underlying laws and design principles –makes the vision a step closer.

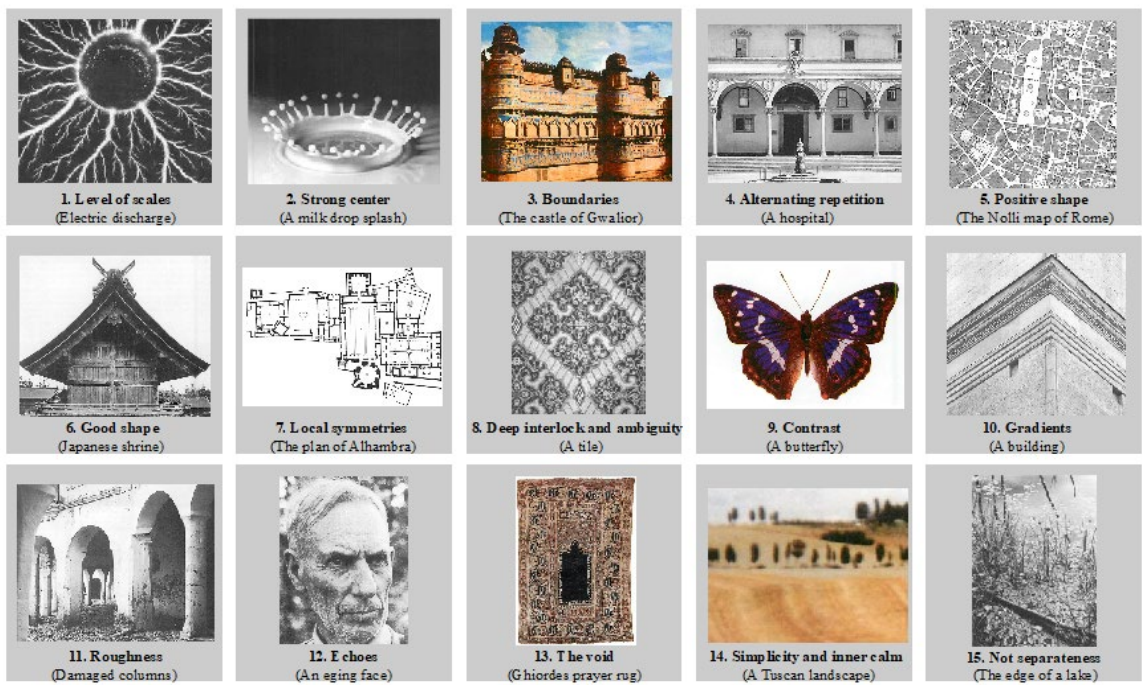

Figure 2: (Color online) The fifteen properties of wholeness (Alexander 2002–2005)
(Note: As a general rule, the degree of wholeness relies on the intensity of the 15 properties: the more intense these properties are, the higher the degree of wholeness, therefore being a living structure.)

In order to see clearly the structure of wholeness, we must adopt the holistic perspective rather than the ordinary analytical perspective to which we have got used. According to Alexander (2002–2005), up to 85% of people see things analytically, so they may not be able to see the wholeness. However, as long as people learn to see things holistically, they will be able to effectively judge degree of wholeness. This is the foundation of the mirror-of-the-self experiment (Alexander 2002–2005, p. 313–350) for objectively judging the degree of wholeness. That is, given two things or two pictures exposed side by side, the human subject is required to pick the one that better reflects him or herself. The experiment is actually to link the external wholeness of the picture to the internal experience of it. Unlike many human tests in psychology that seek an inter-subjective agreement among the subjects, in the mirror-of-the-self experiment, the subject's feeling arises out of the objective wholeness, so that 90% of our feelings are shared.

To further illustrate wholeness or living structure, let us use a spatial configuration consisting of 49 points as a working example (Figure 3). This working example illustrates in a simple manner the mathematical model of wholeness (Jiang 2015c). Suppose that the 49 points have three hierarchical levels – 42 points at the lowest level in blue, one point at the highest level in red, and six points at the middle in green – and that each of the points occupies a piece out of the square space as shown in Figure 3a. Filled with the 49 points, the square space constitutes a spatial configuration in which relationships among individual points or their associated spaces – not only at a same level, but also across different levels – are established, as shown in Figure 3b. Note that relationships at the same level are undirected, while those across different levels are directed from one point of a low level to that of a high level, since usually small things are attracted by large things. Such a spatial configuration can occur within a small-scale space, such as a picture, or a large-scale space, such as 49 cities or human settlements in a vast geographic space. The graph shown in Figure 3b constitutes the structure of wholeness in the context of the square space. Through the graph, the degree of wholeness for individual nodes and for the whole graph can be calculated (Jiang 2015c). Clearly there are far more low degrees of wholeness than high degrees of wholeness; this is largely due to scaling law and Tobler's law discussed in the previous section.

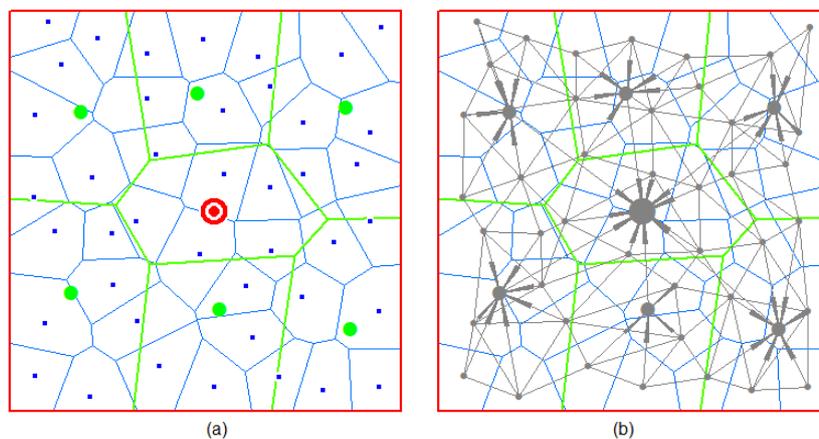

Figure 3: (Color online) Illustration of the concept of wholeness represented a complex network (Note: The wholeness here consists of 49 cities represented by points at three hierarchical levels, indicated by the three colors. The dot sizes to the left show hierarchical levels, while the dot sizes to the right show their degrees of wholeness)

It should be noted that the topological representation of wholeness – as shown in Figure 3 – in which both centers and their relationships provides a general effective framework with which a rich set of social, cultural, ethical, legal, economic, and political relationships may be introduced to make the representation more concrete and comprehensive. For example, one can use administrative or political units and power relationship to construct the wholeness. Although falling outside the scope of this paper, all these different relationships warrant further, promising research in the future.

## 4. Goodness of space as a matter of fact rather than opinion

It is the underlying living structure or wholeness that makes things beautiful or aesthetically pleasing. Goodness of figure, or the law of prägnanz – literally "pregnant with meaning" – or the idea of a "fully developed figure" is probably the most general principle of Gestalt psychology (Koffka 1936). This notion of a "fully developed figure" is always relative; any figure or space or painting – except, of course, master works of art such as Leonardo da Vinci's (1452–1519) *Mona Lisa* and Jackson Pollock's (1912–1956) *Blue Poles* – can be further developed. This goodness of figure is found to depend on some configurational properties (Mowatt 1940), which constitute precursors of the 15 properties of wholeness. In this section, we demonstrate two examples, using the 15 properties and the mathematical model of wholeness (Jiang 2015c), to articulate why one space has a higher degree of goodness, or why one space is objectively more beautiful. The reader can also conduct the mirror-of-the-self experiment to examine if your intuition is consistent with our result, bearing it in mind that you must first learn to see things holistically rather than analytically.

A structure with more centers tends to have a higher degree of goodness than one with few, according to the second property of *"strong centers"*. This is because more centers – if organized well, by which we mean organized complexity (Jacobs 1961, Salingaros 2014) – tend to support each other to constitute a coherent whole or living structure. The first example involves four pairs of figures, as shown in Figure 4. For each of the four pairs, the left has a higher degree of goodness than the right. This can be examined based on the 15 properties. Simply, the left shapes (or small-scale spaces) are able to induce more coherent centers than the right ones. For example, the left shape – shown in pair (a) – has nine centers (the shape itself and eight sides), while the right has five centers (the shape itself and four sides). A circle is a well-shaped form on its own, but not with respect to its surroundings, which look concave. A space is good not only in terms of itself as a closed system, but also according to the larger system that surrounds it. Therefore, a cylindrical column is less beautiful than one with flutes, as fully illustrated in the Temple of Hera at Paestum (Alexander 2002–2005, Volume 1, p. 131–132). This fact is illustrated in pair (b). For the two squares shown in pair (c), the left one, with a tiny dot, is able to induce some 20 centers, while the right (empty square) one has only five centers. More importantly, there is a very steep hierarchy among the 20 centers, but a very flat hierarchy with the five centers (Alexander 2002–2005, Volume 1, p. 81–82, Jiang 2016, p. 479–480). The snowflake (d) is differentiated from the triangle, so the former has a higher degree of goodness than the latter.

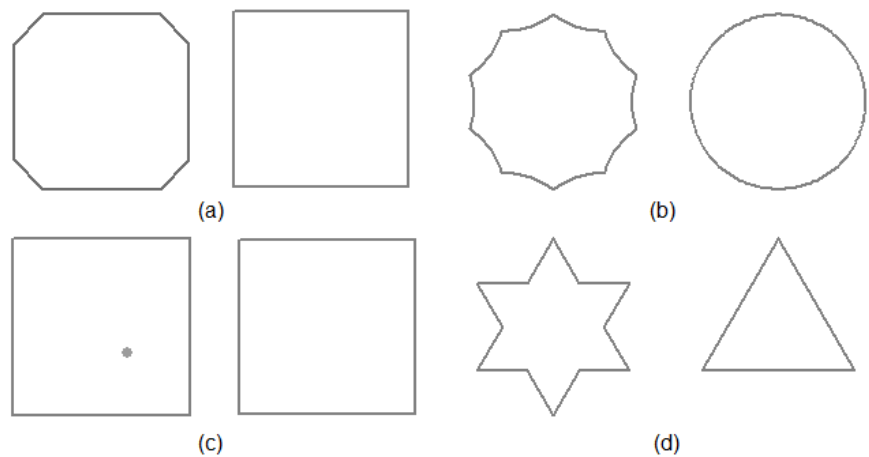

Figure 4: Comparison of goodness of spaces or figures
(Note: For each of the four pairs, the one on the left is further differentiated from that on the right, so the left has a higher degree of goodness.)

The second example involves two spatial configurations that are generated from the 10 numbers 1, 1/2, 1/3, …, 1/10, as discussed in Section 2. The 10 numbers are assigned to random locations in a square space (Figure 5). To some extent, both point patterns in panels (a) and (d) have the same level of differentiation – three hierarchical levels. However, these two patterns differ in terms of adaptation: the points in panel (a) are more adapted than those in panel (d), not only at the same level, but also across different levels. For example, the lowest level in panel (b) is with more or less similar pieces to each other, so it has a high degree of adaptation; this is not the case for the lowest level in panel (e). Across the middle and lowest level in panel (c), the pieces are well adapted each other since each of the green pieces control three or four of the lowest pieces; this is not the case in panel (f). Note that the structures of wholeness are represented as complex networks in panels (d) and (f). The above second example is very much about application of the mathematical model of wholeness (Jiang 2015c). To this point, we can conclude that it is the detailed fine structure – rather than surface geometric shape – that makes things or spaces good. This is exactly why the golden mean – 1.618 – is pervasively seen in beautiful works of art and architecture, not because of the aspect ratio 1.618 superficially, but because of the underlying scaling hierarchy, which organizes complexity to be a living structure (Salingaros 2012). The two examples are just for illustration purpose, and more empirical studies and evidence on living structure can be found in Jiang (2018a) and Jiang and Ren (2018).

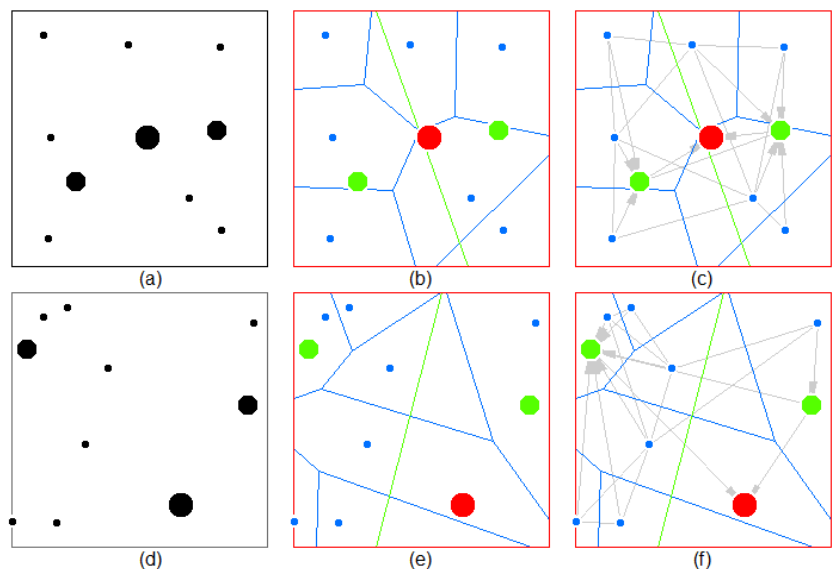

Figure 5: (Color online) Topological representation and analysis of two spatial configurations (Note: The 10 numbers – 1, 1/2, 1/3, …, 1/10 – are assigned to random locations in a square space, generating two different spatial configurations in (a) and (d). The first configuration in (a) has a higher degree of goodness compared to the second in (d). Panels (b) and (e) show how the square spaces are differentiated by the 10 numbers that are put in three hierarchical levels, indicated by three colors. The topological representations shown in panels (c) and (f) indicate that the first configuration (a) is more coherent than the second (d), due to the more balanced topological structure of wholeness for (c) than for (f). The two configurations differ largely in terms of adaptation rather than differentiation.)

Having seen the two examples above, we can make the following remark with respect to the two design principles of differentiation and adaptation. The first example is very much about differentiation, and the good one is more differentiated than the bad one. In other words, the more centers, the more living or more beautiful the space is. The second example is very much about adaptation, for the two spatial configurations, their three levels of scale are already pre-determined. In this case, the adaptation principle plays more important role in determining their goodness. The adaptation is more geometric than statistical, since statistical aspects are already pre-determined by the 10 numbers.

In summary, goodness of space is structural or objective based on the new view of space: space is neither lifeless nor neutral but is a living structure capable of being more living or less living. This new view of space of Alexander differs fundamentally from Newtonian and Leibnizian views of space that are framed under the mechanical world view of Descartes (1637, 1954). Under the new, organic view of space, geography as a science of the Earth's surface has three fundamental issues regarding geographic space: how it looks, how it works, and what it ought to be (Jiang and Ren 2018). The first two issues are largely about understanding the Earth's surface, while the third issue about making and remaking better built environments with a high degree of goodness. The third issue makes geography unique and different from other major sciences, such as physics and biology, since none of these sciences aims for creating living structures that can be objectively measured in terms of their livingness or goodness. It should be noted that this new of space about livingness is shared by many geographers (e.g., Messey 2013), but what is unique for the theory of wholeness is that it provides a mathematical and physical structure for quantifying the livingness in such a powerful way as never before.

**5. Implications of the recursive definition on geography and for sustainability**
The recursive definition of goodness of space has deep implications on geography, and urban design and planning, as well as architecture. All these disciplines deal with goodness of space, although current state of the art of geography largely on description and to a less extent on understanding, while urban design and architecture more concerned with making or creation. If geography – under the new view of space – had the making in its agenda, it would cover urban design and planning and architecture. Fundamentally, geographic space or the Earth's surface in general – according to the new view of space – has the capacity of being more living or less living, more whole or less whole, more sustainable or less sustainable, and even beautiful or ugly. The new view of space implies that – to paraphrase Alexander (1983, as cited in Grabow 1983, p. xi) – geography and architecture would significantly re-shape our world picture of the 21$^{st}$ and 22$^{nd}$ centuries towards an organic one, just as physics did in the 19$^{th}$ and 20$^{th}$ centuries in framing our mechanistic world view. This section further discusses some implications of the recursive definition of goodness of space on geography and for sustainability.

Geography, as a science of the Earth's surface, engages dual aspects about geographic space – understanding and making – but the ultimate goal of geography is to pursue goodness of space or sustainable environments; that is, to make or remake better built environments. This goal – with the focus on making and remaking – makes geography unique and distinct from major sciences such as physics and biology, since the goal of those major sciences has been to understand how things look and work, not – at least not yet – to make or remake things. Of course, design and art are very much about making or remaking, and little about understanding why a work of art or design is beautiful or how much beauty it has. In this regard, the recursive definition offers a definite answer as to not only why a work of art is beautiful, but also how much beauty it has. It is in this circumstance that Alexander (2003) claimed that architecture – or geography, in the context of this paper – is more scientific than other sciences. Alexander's new kind of science – with the focus on the pursuit of beauty or living structure – differs fundamentally from Wolfram's (2003) new kind of science, under which generated patterns are not necessarily living structures. With the generation of living structure as its primary goal, geography will be part of complexity science. However, the mainstream complexity science still focuses on how things look and work rather than how to make beautiful things.

The new view of space adds new insights into understanding of many geographic phenomena as studied in transport geography, or urban geography in general. There is little doubt that the human circulatory system is a living structure that can sustain for a long time without 'traffic jams' or stroke. If our transport systems were planned or designed like the circulatory system or like living structure in general, it would be very smart or sustainable. This way, transport or human activities in general are substantially shaped by the underlying wholeness or living structure. In other words, the complexity of human activities is not due to the complexity of people, but the complexity of the underlying spatial configuration, so-called organized complexity (Jacobs 1961, Salingaros 2014). To better understand human movement or activities, we must first and foremost understand the underlying living structure, and how it shapes our mobility. Also, any new insight into transport or urban geography must converge to underlying living structure, and how the living structure – actually the underlying transport system or urban structure – can be further improved or enhanced towards a higher degree of livingness or goodness.

This paper concentrates essentially on goodness or sustainability of space, with which people have positive feelings such as belonging, well-being or healing. As mentioned at the outset of this paper, this kind of space – or living structure – has another name called place such as homes, workplaces, and neighborhoods to which people belong (Tuan 1977). This human experienced perspective of space or the notion of place is extremely important, since it is people who are the users or inhabitants of space. In this paper, the two concepts of space and place are not separated (Goodchild and Li 2011), but are bridged through the very concept of wholeness. Under the new view of space, goodness of space is not only – first and foremost – defined and measured, being part of physics and mathematics, but also – subsequently – reflected in our minds and cognition, being part of psychology. This is something unique for the new view of space. Through the mirror-of-the-self experiment as mentioned earlier in Sections 3 and 4, the human sense of space can be well examined against the underlying wholeness.

## 6. Conclusion

On the surface this paper presents a recursive definition of goodness of space, but on a deeper level it introduces Alexander's new view of space: space is neither lifeless nor neutral, but a living structure capable of being more living or less living. When geographic space is seen as a living structure, we need to understand how our activities (e.g. movement or traffic) are shaped by the underlying living structure, how our minds reflect the living structure, e.g. the human experienced sense of space, how our well-being is affected by the living structure, and – more importantly – how to make the living structure more living rather than less living, more sustainable rather than less sustainable. These "how" questions require us to go beyond the current mode of thought, mechanical representations of space by geometric primitives, and instead adopt organic representations of individual streets and cities (Jiang 2018a, 2018b) – the so-called topological and scaling perspective – in order to capture the underlying living structure of space. Under the new view of space, there is a shared notion of goodness of space among people regardless of their faiths, ethics, and cultures. It is through the notion of wholeness that the two basic concepts of geography – space and place – get connected, with the latter as human experience of space, coming directly from the former as the structure of wholeness. It is through the notion of wholeness that we can well understand why a space – objectively – has a higher degree of goodness than another.

Now that wholeness indeed exists in our surroundings, research on urban structure and dynamics for sustainability must closely connect to the underlying structure of wholeness, and how it can be improved, enhanced or reinforced. Now that wholeness can indeed serve as a scientific foundation, geography or architecture will be significantly reshaped towards a major science, becoming part of complexity science not only for understanding complexity, but also for creating complex and living structures. This vision with a focus on creation, as proposed initially for architecture by Alexander (2003), has already exceeded the current state of the art of complexity science that is largely constrained on an understanding of complexity. Unlike major sciences, such as physics and biology, geography – as now conceived under the new view of space – is inevitably able to touch on the fundamental issue of design, with the aim of making better, smarter and sustainable built environments.

**Acknowledgement**
XXXXXXXXX